\documentclass[prl,twocolumn,showpacs,superscriptaddress]{revtex4}

\def\H{H}
\def\A{A}
\def\B{B}

\def\x{X}

\def\z{Z}

\usepackage{subfigure}
\usepackage{amsfonts}
\usepackage{graphicx}
\usepackage{color}

\begin{document}

\title{Entropic Barriers for Two-Dimensional Quantum Memories}

\author{Benjamin J. Brown}
\email{benjamin.brown09@imperial.ac.uk}
\affiliation{Quantum Optics and Laser Science, Blackett Laboratory, Imperial College London, Prince Consort Road, London, SW7 2AZ, United Kingdom}
\affiliation{School of Physics and Astronomy, University of Leeds, Leeds, LS2 9JT, United Kingdom}
\author{Abbas Al-Shimary}
\affiliation{School of Physics and Astronomy, University of Leeds, Leeds, LS2 9JT, United Kingdom}

\author{Jiannis K. Pachos}
\affiliation{School of Physics and Astronomy, University of Leeds, Leeds, LS2 9JT, United Kingdom}



\begin{abstract}
Comprehensive no-go theorems show that information encoded over local two-dimensional topologically ordered systems cannot support macroscopic energy barriers, and hence will not maintain stable quantum information at finite temperatures for macroscopic timescales. However, it is still well motivated to study low-dimensional quantum memories due to their experimental amenability. Here we introduce a grid of defect lines to Kitaev's quantum double model where different anyonic excitations carry different masses. This setting produces a complex energy landscape which entropically suppresses the diffusion of excitations that cause logical errors. We show numerically that entropically suppressed errors give rise to super-exponential inverse temperature scaling and polynomial system size scaling for small system sizes over a low-temperature regime. Curiously, these entropic effects are not present below a certain low temperature. We show that we can vary the system to modify this bound and potentially extend the described effects to zero temperature.
\end{abstract}

\maketitle


It is not well understood what properties a locally interacting system needs to possess to maintain encoded quantum information at finite temperatures \cite{DennisKitaevLandahlPreskill,Bacon06}. In the presence of a finite temperature information encoded in the ground space of a gapped system will eventually decohere. This occurs when {\em logical} errors happen with high probability. Assuming that the thermal generation of errors follows Arrhenius' law the corresponding coherence time is of the order of
\begin{equation}
\tau \sim e^{\beta \epsilon},
\label{eqn:Arrhenius}
\end{equation}
where $\epsilon$ is the energy cost of typically occurring errors that decohere information and $\beta = 1/T$ is the inverse temperature. 

Topological systems \cite{wenmanybody, Pachos} are popular platforms for quantum memories. They are resilient against local perturbations that can cause coherent errors~\cite{WenNiu90, BravyiHastingsMichalakis10}. They can also be easily modified to improve their resilience against probabilistic errors \cite{WoottonPachos, Stark}. Ideally, we would like the coherence time of such systems to grow efficiently as we increase their linear size $L$ at non-zero temperature. No-go theorems \cite{NussinovOrtiz07, BravyiTerhal09, LandonCardinalPoulin} show that two-dimensional topologically ordered Hamiltonians with local interactions must have a constant energy gap, $\Delta$, between two ground states. For such systems we expect $\epsilon \propto \Delta $ and from (\ref{eqn:Arrhenius}) we obtain a coherence time that is independent of $L$ \cite{AlickiFannesHorodecki, ViyuelaRivasMartin-Delagado}.

Pioneering work with three-dimensional models has shown the realisation of {\em fragile glassy} quantum systems~\cite{CastelnovoChamon11, BravyiHaah13}. These systems have local error excitations created with energy penalty $\Delta$, while they can thermally increase their size $\xi$ with a logarithmic energy cost $\epsilon \sim \kappa \Delta  \ln \xi $ for some constant $\kappa > 0$ \cite{GarrahanNewman, BravyiHaah11, CastelnovoChamon11}. Given the density of local excitations scales exponentially with inverse temperature, $e^{\beta \Delta}$, typical excitations incident to the lattice have an average separation $\mathcal{R} \sim e^{\beta \Delta / d}$, where $d$ is the dimensionality of the system. Assuming that the memory decoheres when the size $\xi$ of the errors becomes comparable to $ \mathcal{R}$, we find that the logical errors have an energy cost that grows linearly with inverse temperature, $\epsilon \sim \beta \Delta\kappa/d$. From (\ref{eqn:Arrhenius}) we find that such systems have coherence times which scale super-exponentially with inverse temperature
\begin{equation}
\tau \sim e^{ \beta^2 \kappa \Delta^2  / d}.
\label{eqn:super}
\end{equation}  
This also gives rise to polynomial system-size scaling in the limit of small system sizes. Indeed, at low temperatures where $\mathcal{R} \sim L$ we obtain
\begin{equation}
\tau \sim L^{  \beta \kappa \Delta},
\label{eqn:size}
\end{equation}
where $L$ is smaller than a critical size $L^* $ which increases exponentially with $\beta$ \cite{BravyiHaah13}. Such behaviour was coined {\em partial self correction} in Ref.~\cite{BravyiHaah13}.

Here we take an alternative approach to two-dimensional topological memories. We consider systems where excitations are {\em entropically steered} into high-energy configurations. 
In particular, we introduce a {\em grid of defects} \cite{BeigiShorWhalen, KitaevKong, BarkeshliJianQi} to a topologically ordered model. For such a model we take Kitaev's $\mathbb{Z}_N$ quantum double with $N>2$ \cite{Kitaev03}, designed so that its anyonic excitations have differing masses. This imbalance, together with the presence of the grid, and the splitting structure of the anyon model, creates a complex energy landscape for errors diffusing across the lattice. We numerically demonstrate that this system exhibits super-exponential coherence scaling (\ref{eqn:super}) with $\beta$ in a low temperature regime, as well as coherence time scaling of the form of (\ref{eqn:size}) in the limit of small system sizes. We observe that the super-exponential behaviour is witnessed only for a finite range of temperatures that is tuneable by the grid spacing. This is because the entropic effect is thermally activated. Our model is a first step for combating temperature errors with two-dimensional topological models that are experimentally amenable using existing technologies \cite{Gladchenko}.

\begin{figure}
\includegraphics[scale = 1.5]{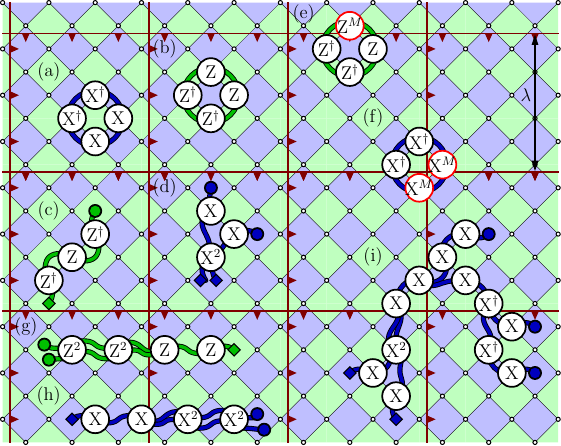}
\caption{Kitaev's $\mathbb{Z}_N$ model on an $L\times L$ periodic lattice modified by an oriented defect grid. $N$-dimensional spins reside at the vertices of the lattice. (a) The primal $A_v$ and (b) the dual $B_p$ operators of the Hamiltonian. (c,d) String-like operators with anyonic excitations at their end-points. Modified dual (e) and primal (f) operators placed along the defect lines. (g,h) The charge of anyonic excitation crossing a defect line is multiplied by $M$. (i) A possible string operator with multiple end-points. The energy or ``mass" of such an operator increases with the number of anyons it creates, i.e. with the number of its end-points. \label{stabilisers}}
\end{figure}

We begin the discussion by reviewing Kitaev's $\mathbb{Z}_N$ quantum double model \cite{Kitaev03}. Consider $N$-level spins on the vertices of a diamond lattice, as shown in Fig.~\ref{stabilisers}. The faces of the lattice are bicolored as primal (green) faces, $v$, and dual (blue) faces, $p$. We define the system size by arranging primal faces on an $L \times L$ periodic square lattice.
 The spins interact by mutually commuting local projectors $ P^a_v$ and $ Q^a_p $ for each $v$ and $p$ respectively, where $a \in 1,2,\dots, N-1$. These projectors are given by $P^a_v = \frac{1}{N}\sum_{k=1}^N e^{2 ak \pi i /N} \A_v^k$ and $Q^a_p = \frac{1}{N}\sum_{k=1}^N e^{2 a k \pi i/N} \B_p^k$, where $\A_v \, ( \B_p)$  are known as primal\,(dual) operators, shown in Fig.~\ref{stabilisers}(a) and (b), respectively. Operators $\x$ and $\z$ are generalised Pauli matrices with $\x | k \rangle = | k+1( \text{mod}\, N) \rangle$ and $\z |k \rangle = e^{2  \pi k i  / N} | k \rangle$.
We then write the Hamiltonian
\begin{equation}
\H =  \sum_v {\bf J} \cdot {\bf P}_v + \sum_p {\bf J} \cdot {\bf Q}_p,
\label{eqn:Ham1}
\end{equation}
where ${\bf J} =  (J_1, J_2 , \dots , J_{N-1})$ are interaction strengths, ${\bf P}_v = (P^1_v, P^2_v , \dots , P_v^{N-1})$ and ${\bf Q}_p = (Q_p^1, Q_p^2 , \dots , Q_p^{N-1})$. The model has a $N^2$-fold degeneracy, which we use for the logical encoding space of the memory. 

The excitations of Hamiltonian (\ref{eqn:Ham1}) are {\em anyonic} \cite{Pachos}. There are two anyonic flavours: electric anyons, $e_k$, and magnetic anyons, $m_k$, which carry charge $k  =  1, \dots, N-1$. The $e_k$ and $m_k$ excitations live exclusively on the primal and dual faces, respectively. In general, these anyons have different masses depending on the choice of ${\bf J}$. In fact, experimental architectures that may realise such a Hamiltonian are likely to have such an imbalance in masses \cite{WoottonLahtinenDoucotPachos}. In this work we consider particular mass imbalances which embellish the effects we demonstrate. The eigenvalues $E = \sum_k (n_{m_k} + n_{e_k}) J_k $ of the Hamiltonian increase with the number of anyons $n_a$ of type $a$ that emerge as localised excitations. Examples of string operators that create anyons at their end points are shown in Fig.~\ref{stabilisers}(c) and (d). String operators in this model for $N>2$ can split, as shown in Fig.~\ref{stabilisers}(d). The allowed splitting channels are governed by the fusion rules that preserve the total charge modulo $N$ \cite{Kitaev03}. 

We next consider the time evolution of the a lattice subject to an Ohmic heat bath. We model the bath by the spin-boson noise model in the Davies weak coupling limit. The thermal evolution of the lattice is determined by the rate equation \cite{AlickiFannesHorodecki, ChesiRothlisbergerLoss}
\begin{equation}
\gamma( \omega ) =  { \omega}/({1-e^{-\beta \omega}}),
\label{eqn:rate}
\end{equation}
where $\beta=1/T$ is the inverse temperature of the model and $\omega$ is the energy change due to the error operation. As this rate satisfies the detailed balance equation, it eventually leads to a thermal state $\rho \sim e^{-\beta H}$. The induced phase-flip and bit-flip errors have equivalent effects on the lattice. As they can be considered independently of one another we model only $\x^a_k$ type errors.

We simulate the $e_k$ excitations of the $\mathbb{Z}_5$-quantum double model in a thermal environment using classical random walks. We define the coherence time $\tau$ to be the time where the probability $p$ of recovering the encoded ground state falls below $0.99$. We evaluate $p$ using $4\cdot10^4$ Monte Carlo samples. For a single sample we encode the lattice in a specific ground state, and then simulate the lattice dynamics with respect to rate Eqn.~(\ref{eqn:rate}) with inverse temperature $\beta$. We then attempt to recover the encoded information using a decoding algorithm. A decoder uses the positions of excitations on the lattice to predict the most likely error configuration consistent with the given anyonic configuration to attempt to recover the initially encoded ground state. We use the {\em HDRG decoder} described in Ref.~\cite{Anwar}. We do not attempt to decode errors larger than $L/3$, as this reduces finite size effects when considering small lattices \cite{BravyiHaah13}. 

We identify the diffusive nature of excitations in two-dimensional systems in the low-temperature limit by evaluating coherence times for Hamiltonian~(\ref{eqn:Ham1}) where $J_1 = J_4 \equiv J_L = 0.38$ and  $J_2 = J_3 \equiv J_H = 1.0$. We examine the low temperature regime $ 6 \lesssim \beta \lesssim 9 $ where excitations $e_1$ and $e_4$ are created most commonly in pairs with $\Delta = 0.76$. In this temperature regime we find $\tau \sim e^{ 0.53 \beta - 2.5}$. We observe $\epsilon = 0.53$ where remarkably $ \epsilon <  \Delta$. This discrepancy can be explained by considering the propagation of excitations. While errors are created at a rate exponentially suppressed by the Hamiltonian gap $ \gamma_{\text{crtn.}} \sim e^{- \beta \Delta}$, the initially created excitations can generate logical errors by thermal propagation, which occurs with rate $\gamma_{\text{prgtn.}}\sim 1/\beta $. As $\gamma_{\text{prgtn.}} / \gamma_{\text{crtn.}}  $ diverges with $\beta$, the effect of errors caused by propagation become more appreciable as we decrease the temperature, thus resulting to an effective energy $\epsilon$, which is smaller than the local error generation cost $\Delta$.

To reduce the long range propagation of excitations at low temperatures we introduce charge $M$-modifying defect lines to the topological model. These lines map crossing excitations to excitations with different anyonic charge $e_k \rightarrow e_{l}$ and $m_{l}\rightarrow m_k$. The defect lines are introduced in Hamiltonian (\ref{eqn:Ham1}) by modifying only the $P^a_v$ and $Q^a_p$ terms that lie along the red lines shown in Fig.~\ref{stabilisers}. We frame one side of the line which we mark with triangles. Along these lines we replace the primal and dual operators of projectors $P^a_v$ and $Q^a_p$ with new $\A_v$ and $\B_p$ operators which are changed as follows: if a defect line lies on primal\,(dual) face $v\,(p)$, then we raise the support of the $\A_v\,(\B_p)$ operator the framed\,(unframed) side of the defect line to the power $M$. We show explicit examples of modified $\A_v$ and $\B_p$ operators in Fig.~\ref{stabilisers}(e) and (f).  The $M$-modifying defect lines affect the $e_k$ ($m_k$) anyons crossing the line in the negative (positive) direction by multiplying their charge by M mod$\,N$, as shown in  Fig.~\ref{stabilisers}(g) and (h), respectively, for $M=2$. The inverse operation occurs if an anyon crosses in the opposite direction. Here we take a regular defect grid of separation $\lambda$, as shown in Fig.~\ref{stabilisers}. 

For concreteness we consider a grid of $M=2$ defects to the simulated $\mathbb{Z}_5$ model. In this case propagating low mass excitations, $e_1$ and $e_4$, across the defect lines will morph into excitations of higher mass,  $e_2$ and $e_3$, and are thus energetically penalised by $J_H - J_L$. To see how a defect grid encourages excitations to grow in mass with size $\xi$ we analyse the following common process. Consider the single high-mass excitation which has arisen from a low-mass excitation hopping a defect line. This excitation can continue to propagate across a subsequent defect line unimpeded. However, we choose $J_L / J_H$ and $\lambda$ such that the high mass excitations decay rapidly into two low mass excitations via the fusion channels of the anyon model. This decay process is mediated by the heat bath. The propagation of the new pair of low mass excitations is further impeded by their enclosing defect lines. If these excitations further propagate through the grid then their mass is expected to further increase by $J_H - J_L$. Hence, the total energy will increase as a function of $\xi$ by the above mechanism. An example of a typical evolution is shown in Fig.~\ref{stabilisers}(i). Nevertheless, it is possible, though unlikely, that the high mass excitation propagates without decay and crosses defect lines without any energy penalty. This process can be suppressed by choosing the distance between defect lines, $\lambda$, large.

\begin{figure}
\includegraphics[width=\columnwidth]{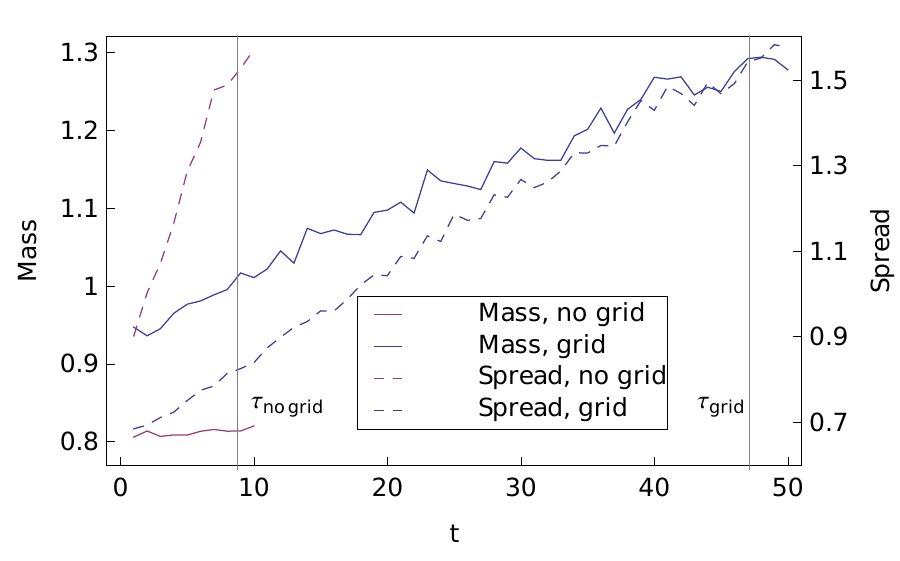}
\caption{ The time evolution of an initial single pair of excitations subject to a thermal environment. We modify the environment so that excitations may only propagate, split and fuse. No new excitations can be created. The average total mass (solid lines) and average spread (dashed lines) of the evolution are depicted without (purple) and with (blue) a defect grid until thermalisation. In the absence of the grid the number of excitations remains constant, while they spread rapidly. In the presence of the grid the propagation of excitations is accompanied by a continuous increase in mass. At the same time the excitations spread with a slower velocity. The depicted thermalisation times, $\tau_\mathrm{no \,\,grid}$ and $\tau_\mathrm{grid}$, correspond to the complete thermal evolution of the system. The evolutions are obtained for $\lambda=2$, $\beta = 8$, $J_L = 0.4$ on an $L=48$ lattice. We average over one thousand samples.
\label{fig:twofortea}}
\end{figure}

To identify the new dynamics introduced by the defect line, we first study a single excitation on a lattice in Fig.~\ref{fig:twofortea}. This is well motivated in the low temperature regime where excitations are sparsely created. We study the mass of a single error which initially generates a pair of low mass excitations for a lattice with  $J_L = 0.4$, $J_H = 1.0$, $\lambda = 2$ and $\beta = 8.0$. We monitor the increase in mass due to the propagation of these excitations through the defect grid until the coherence time, $\tau$, of the system. The solid blue line in Fig.~\ref{fig:twofortea} shows that the mass of the error grows as it diffuses over time. This contrasts from a single excitation on a lattice with no defects which does not appreciably increase in mass, shown by the solid purple line in Fig.~\ref{fig:twofortea}. We consider also the rate at which errors diffuse over the lattice. We measure diffusion, by measuring average distance the excitation mass has travelled from the centre of mass of all the excitations on the lattice. We see that the energy barriers we introduce substantially slow the rate that excitations spread across the lattice, shown by the dotted blue and purple lines of Fig.~\ref{fig:twofortea}. Hence, the presence of the defect grid significantly alters the dynamics involved in the propagation of excitations causing a dependence of the total energy on their spread $\xi$. 

\begin{figure}
\includegraphics[width=\columnwidth]{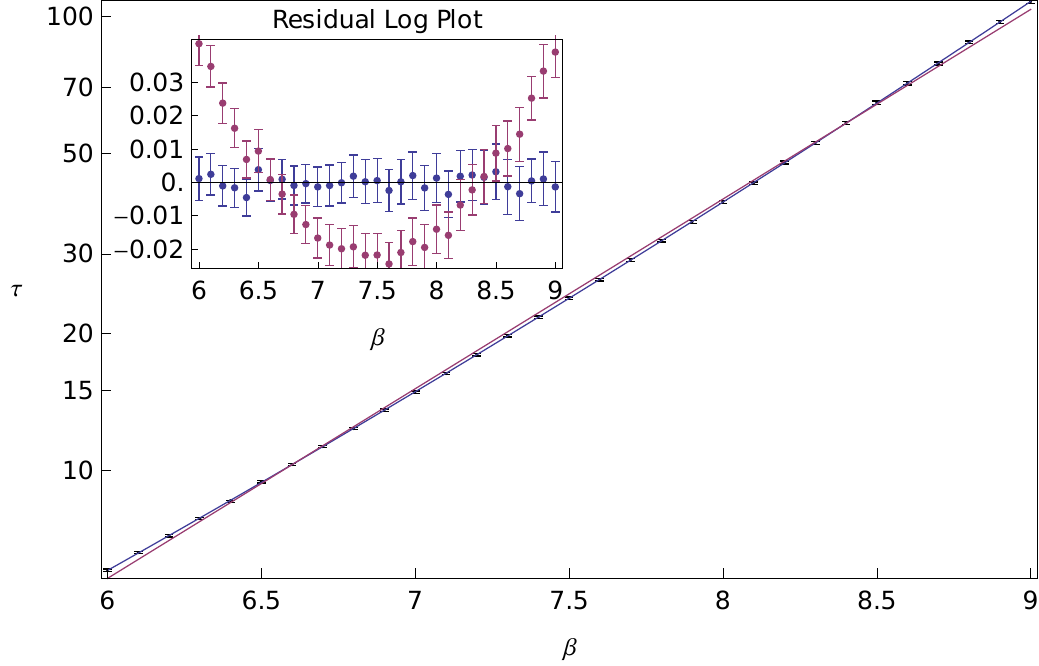}
\caption{ \label{fig:timeaftertime} The coherence time, $\tau$, of the system with defect grid as a function of $\beta$. We take $L=72$ and $J_L/J_H = 0.38$. $\lambda$ alternates between 1 and 2 along the lattice in both directions so that a variety of system sizes can be probed. Inset depicts the residual logarithmic plot of $\tau$. The purple points correspond to the linear fitting and the blue to the quadratic fitting. A super-exponential behaviour is obtained with fitting $\tau \sim \exp(0.028 \beta^2 + 0.54\beta-2.5)$ to be compared with (\ref{eqn:super}). The linear fit gives $\tau \sim \exp (0.96\beta - 4.0)$.}
\end{figure}

\begin{figure}[t]
\includegraphics[width=\columnwidth]{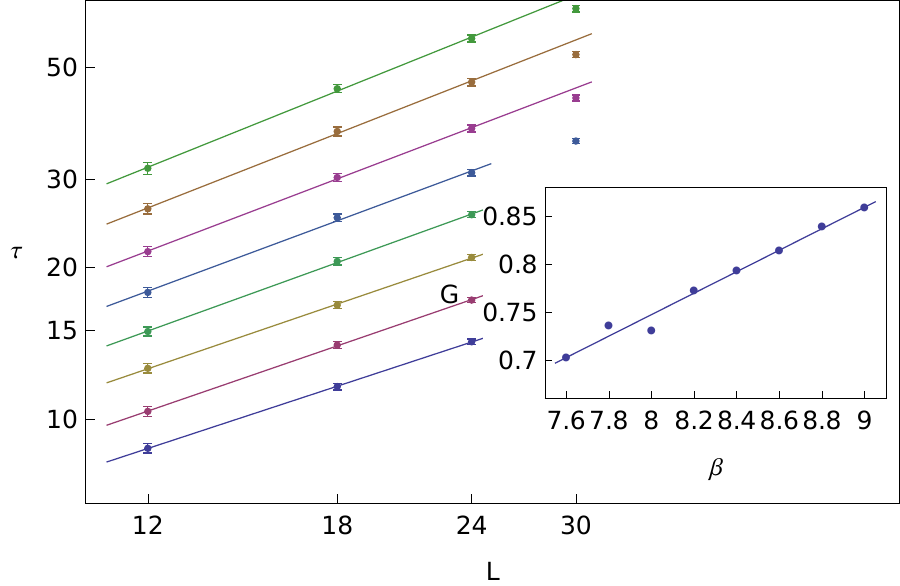}
\caption{\label{Fig:SystemSizes} The coherence time as a function of $L$, using the same parameters as in Fig. \ref{fig:timeaftertime}. Fitting lines are displayed for $\beta = 7.6,7.8, \dots, 9.0$ ordered from bottom to top. Inset depicts the gradients $G$ of the fitting for different temperatures, showing a polynomial degree that increases linearly with $\beta$. Fitting gives $\tau \sim L^{0.11 \beta - 0.15}$ to be compared with (\ref{eqn:size}).}
\end{figure}

We now evaluate coherence times of a lattice with defect lines where $J_L = 0.38$ and $J_H = 1.0$. We consider a lattice where we alternate defect line separation between $\lambda = 1$ and $\lambda = 2$ along the lattice. This does not provide optimal results, but it enables us to probe small system sizes without affecting the ground state degeneracy of the lattice. The resulting coherence time as a function of $\beta$ is shown in Fig.~\ref{fig:timeaftertime} where $L=72$. The data are well fit by the function $\tau \sim e^{0.028\beta^2 + O(\beta)}$. Hence, we identify weak super-exponential inverse temperature scaling. We also find polynomial system size scaling for small system sizes, where the degree of the polynomial scales linearly with $\beta$, such that $\tau \sim L^{0.11\beta +O(1)}$, shown in Fig.~\ref{Fig:SystemSizes}. We can compare these fittings with equation (\ref{eqn:super}) and (\ref{eqn:size}), respectively, to find $\Delta \sim 0.5$ and $\kappa \sim 0.2$. 
The small value of  $\kappa$ compared with the cubic code values~\cite{BravyiHaah13} is due to the small rate at which new excitations are created with $\xi$ in our model. 

Curiously, we do not experience super-exponential inverse-temperature scaling in the $\beta \rightarrow \infty$ limit. If the temperature becomes too low, low-mass excitations will typically propagate by recombining with one another and move across defect lines as high mass excitations with no energy penalty. For fixed $\lambda$ this process occurs exponentially more frequently than single low mass excitations propagating over defect lines as inverse temperature increases. This is because recombination occurs at a lower energy cost, $J_H - 2J_L$, than the process of a low-mass excitation propagating across a defect line, which costs $J_H - J_L$. If this occurs readily we loose the super-exponential inverse temperature scaling. We therefore require a suitably warm low-temperature environment to observe the effects we describe. To compensate this effect we increase $\lambda$ in order to reduce the rate of excitation recombination. We find for $\lambda = 6$ on $L=96$ lattice a coherence time $\tau = \exp(0.044\beta^2 - 0.29\beta + 2.2)$ in the significantly lower temperature regime $13 \lesssim \beta \lesssim 16$. Ideally we would like to simulate very low temperatures with large grid spacing, $1 \ll \lambda \ll \mathcal{R}$, where high-mass excitations decay with near certainty and thus optimise $\kappa$. In the present case we see that for $\beta \sim 16$ we have $\epsilon \sim 1.13$, which is greater than the studied $\lambda =1,\,2$ case, where we find an optimal $\epsilon \sim 1.04$ at $\beta \sim 9$. This is suggestive that we may extend the described effects to lower temperatures by exploring a wider parameter space. It will be interesting if we can amplify the entropic effects we present here to find a parameter regime where $\epsilon \gg \Delta$. We leave a comprehensive study of varying $\lambda$ for future work.

In this letter we have presented a two-dimensional topologically ordered memory that shows weak super-exponential inverse temperature scaling, and polynomial system size scaling for small system sizes in a low temperature regime. Remarkably, we achieve these effects by introducing charge modifying defect lines to the considered model which entropically steer excitation dynamics into high energy configurations. We notice also that, due to recombination events, we do not observe these properties at very low temperatures. We have presented preliminary results showing that we can overcome this by separating the defect grid as we lower the temperature.

We acknowledge S. Barrett, C. Brell and J. Wootton for inspiring discussions. We also thank H. Anwar, S. Burton and G. Duclos-Cianci for conversations about decoding algorithms. Computational resources were provided by the Imperial College High Performance Computing Service. This work is supported by the EPSRC.

\bibstyle{plain}

\end{document}